\documentstyle[12pt]{article}
\def\lsim{\mathrel{\lower2.5pt\vbox{\lineskip=0pt\baselineskip=0pt
           \hbox{$<$}\hbox{$\sim$}}}}
\def\gsim{\mathrel{\lower2.5pt\vbox{\lineskip=0pt\baselineskip=0pt
           \hbox{$>$}\hbox{$\sim$}}}}
\begin{document}
\setlength{\baselineskip}{8mm}
\setcounter{page}{1}
\thispagestyle{empty}
\begin{center}
{\large  \bf
Quark-Lepton Masses and the CKM Matrix \\
in a String Inspired Model }
\end{center}
\vskip 10mm
\begin{center}
Naoyuki HABA$^1$, Chuichiro HATTORI$^2$, \\
Masahisa MATSUDA$^3$ and Takeo MATSUOKA$^1$  \\
{\it
${}^1$Department of Physics, Nagoya University \\
           Nagoya, JAPAN 464-01 \\
${}^2$Science Division, General Education \\
     Aichi Institute of Technology \\
      Toyota, Aichi, JAPAN 470-03 \\
${}^3$Department of Physics and Astronomy \\
     Aichi University of Education \\
      Kariya, Aichi, JAPAN 448 \\
}
\end{center}
\vspace{10mm}
\begin{abstract}
In the context of the Calabi-Yau string model
we discuss the origin of characteristic pattern
of quark-lepton masses and the CKM matrix.
The discrete $R$-symmetry $Z_{2k+1} \times Z_2$
is introduced and
the $Z_2$ is assigned to the $R$-parity.
The gauge symmetry at the string scale
is broken into the standard model gauge group
at an intermediate energy scale.
At energies below the intermediate scale
down-type quarks and also leptons are mixed
with unobserved heavy states, respectively.
On the other hand, there are no such mixings
for up-type quarks.
The large mixings between light states and heavy ones
cause different mass pattern of leptons and
down-type quarks from that of up-type quarks
and yield a nontrivial CKM matrix for quarks
but a trivial one for leptons.
The mixing mechanism is illustrated using
a two-generation toy model.
\end{abstract}
\newpage
\section{Introduction}
\hspace*{\parindent}
The characteristic pattern of quark-lepton masses
has long been a challenging problem to explain
its origin.
With the aim of solving this problem
it is interesting to explore
how Planck scale physics (superstring theory)
determines low-energy parameters in
the effective theory.
The observed masses of quarks and leptons have
the hierarchical pattern
\begin{description}
\item [(i)] \qquad \qquad $m$(1\,st gen.)
          $\ll $ $m$(2\,nd gen.) $\ll $ $m$(3\,rd gen.)
\end{description}
and also the ratios are in line as
\begin{description}
\item [(ii)] \qquad \qquad $m_u/m_d
                 < m_c/m_s < m_t/m_b$.
\end{description}
Up to now several possibilities of explaining
these features have been studied by many authors
\cite{Bound}
\cite{SO10}
\cite{Nambu}
\cite{Infra}.
A possibility is that all the observed pattern
of fermion masses are attributable to
the boundary condition,
i.e. to the hierarchical structure of Yukawa
couplings themselves at a very large scale.
However, when we take GUT-type models,
it is difficult to find a satisfactory solution
in which property (ii) comes into line with
a simple unification of Yukawa couplings.
In this paper we explore a somewhat distinct possibility
and propose a new type of model
which potentially generates the characteristic
pattern of fermion masses and
the Cabibbo-Kobayashi-Maskawa(CKM) matrix.
In the model property (i) is attributed to
the texture of renormalizable and nonrenormalizable
interactions restricted by discrete symmetries
at the string scale.
This mechanism is similar to that
proposed in Ref.\cite{Bound}.
On the other hand, property (ii) comes from
large mixings among states observed at low energies
and unobserved heavy ones.
The mixings occur only for down-type quarks and
for leptons below the energy scale at which
the gauge group is broken into
the standard model gauge group
$G_{st} = SU(3)_c \times SU(2)_L \times U(1)_Y$.
Our study is made in the context of Calabi-Yau
string model with Kac-Moody level-one.

In the four-dimensional effective theory from
Calabi-Yau string compactification
there are many peculiar features beyond the minimal
supersymmetric standard model (MSSM).
First point is that the gauge group $G$,
which is given via the flux breaking at
the string scale,
would be larger than the standard model
gauge group $G_{st}$.
As a simple example we will choose
$G = SU(6) \times SU(2)_R$,
under which doublet Higgs and color-triplet
Higgs fields transform differently
\cite{Aligned}.
As we will see later, the gauge group $G$ is
spontaneously broken to $G_{st}$ in two steps
at intermediate energy scales.
Second point is that
the massless sector of the Calabi-Yau string model
contains particles beyond the ones in the MSSM.
In string inspired models we typically have
a number of generations and anti-generations.
For illustration, if the gauge group $G$ is $E_6$,
the massless chiral superfields apart from
$E_6$-singlets consist of
\begin{equation}
       N_f \,{\bf 27} \
               + \ \delta \,({\bf 27} + {\bf 27^*})\,,
\end{equation}
where $N_f$ means the family number at low energies.
It should be noted that $\delta $ sets of
vector-like multiplets are included in
the massless sector.
We will assume $\delta = 1$ for simplicity.
In addition, particles beyond the MSSM are
contained also in ${\bf 27}$-representation of $E_6$.
Namely in ${\bf 27}$ we have
quark superfields $Q = (U, D)$, $U^c$, $D^c$,
lepton superfields $L = (N, E)$, $N^c$, $E^c$,
Higgs doublets $H_u$, $H_d$,
color-triplet Higgses $g$, $g^c$ and
an $SO(10)$-singlet $S$.
When the gauge group $G$ is broken into $G_{st}$,
superfields $D^c$ and $g^c$ ($L$ and $H_d$) are
indistinguishable from each other under $G_{st}$.
Hence there possibly appear mixings between $D^c$
and $g^c$ and between $L$ and $H_d$.
On the other hand, for up-type quarks there appear
no such mixings.
Even if up-type, down-type quarks and leptons
share their interactions in common at $M_S$,
$D^c$-$g^c$ mixing and $L$-$H_d$ mixing potentially
cause different mass pattern of down-type quarks
and leptons from that of up-type quarks.
Further the mixings may be responsible
for the CKM matrix.
Third point of peculiar features beyond the MSSM is
that superstring theory naturally provides
the discrete symmetries
which stem from symmetric structure of the
compactified space.
As shown in Gepner model
\cite{Gepner},
the discrete symmetry can be the $R$-symmetry
under which the components of a given superfield
transform differently.
Also, the discrete symmetry could be used
as a horizontal symmetry.
The discrete $R$-symmetry strongly limits
the renormalizable and nonrenormalizable interactions
and then possibly controls parameters in the
low-energy effective theory.
Recently it has been argued that the discrete
$R$-symmetry controls energy scales of
the symmetry breaking
\cite{Disc},
the magnitude of Majorana masses of
the right-handed neutrino
\cite{Majorana}
and the stability of the weak-scale hierarchy
\cite{Tadpole}.
We will introduce the discrete $R$-symmetry
$Z_{2k+1} \times Z_2$ at the string scale.

In this paper we explore how both of
the mixing mechanism mentioned above
and the discrete symmetry generate
the fermion mass pattern and the CKM matrix.
This paper is organized as follows.
In section 2 we introduce the discrete $R$-symmetry,
which puts stringent constraints on
interactions in the superpotential.
The $Z_2$ symmetry is chosen so as to be in
accord with the so-called $R$-parity in the MSSM.
We assume that $R$-parity is conserved
over the whole energy range from the string scale
to the electroweak scale.
After arguing that the discrete $R$-symmetry
controls energy scales of the gauge
symmetry breaking,
we study particle spectra of vector-like
multiplets in section 3.
Since doublet Higgses and color-triplet Higgses
belong to the different representations of
the gauge group $G = SU(6) \times SU(2)_R$,
we can find solutions in which color-triplet Higgs
mediating the proton decay are sufficiently heavy
while keeping a set of Higgs doublets light
without some fine-tuning of parameters.
In section 4 mass matrices for chiral multiplets
are presented.
By using the perturbative method
mass eigen-equations are derived.
Mixings between $D^c$ and $g^c$ are also discussed.
Choosing appropriate assignments of discrete charges,
we get large mixings between them.
Due to this mixing the mass pattern of
down-type quarks is possibly changed from
that of up-type quarks.
The model potentially generates not only
the hierarchical pattern of quark masses
but also the texture of the CKM matrix.
In section 5 we discuss mixings between $L$ and
$H_d$ and study spectra of leptons.
The CKM matrix for leptons turns out to be
a unit matrix.
To illustrate the general features of the mixing
mechanism,
we exhibit a two-generation toy model in section 6.
In the final section we conclude with a brief
summary of our results.


\section{Discrete $R$-symmetry}
\hspace*{\parindent}
In order to guarantee the stability of
the weak-scale hierarchy without fine-tuning,
it is favorable that doublet Higgses and color-triplet
Higgses reside in different irreducible
representations of the gauge group $G$
derived via the flux breaking at the
string scale.
As the largest gauge group implementing such
a situation is $G = SU(6) \times SU(2)_R$
\cite{Aligned},
we choose $SU(6) \times SU(2)_R$
as an example of $G$ in this paper.
Main points obtained remain available also
in the other cases of $G$ under which
doublet Higgses and color-triplet Higgses
transform differently.
Chiral superfields $(\Phi )$ in ${\bf 27}$
representation of $E_6$ are decomposed into
\begin{eqnarray*}
\Phi ({\bf 15, 1})  &:& \ \ Q, L, g, g^c, S, \\
\Phi ({\bf 6^*, 2}) &:& \ \ U^c, D^c, N^c, E^c,
                                      H_u, H_d.
\end{eqnarray*}
Although $L$ and $H_d$ ($D^c$ and $g^c$) have
the same quantum numbers under $G_{st}$,
$L$ and $H_d$ ($D^c$ and $g^c$) belong to
different irreducible representations of
$SU(6) \times SU(2)_R$.
The superpotential $W$ is described in terms of
${\bf 27}$ chiral superfields $(\Phi )$ and
${\bf 27^*}$ ones $(\overline {\Phi })$ as
\begin{equation}
     W = \Phi ^3 + \overline {\Phi}^3 +
            (\Phi \overline {\Phi })^{m+1} +
                \Phi ^3 (\Phi \overline {\Phi })^n
                   + \cdots ,
\end{equation}
where $m$ and $n$ are positive integers and
all the terms are characterized by
the couplings of $O(1)$ in
$M_S = O(10^{18}{\rm GeV})$ units.
The cubic term $\Phi ^3$ is of the forms
\begin{eqnarray}
    (\Phi ({\bf 15, 1}))^3 & = & QQg + Qg^cL + g^cgS, \\
    \Phi ({\bf 15, 1})(\Phi ({\bf 6^*, 2}))^2 &
            = & QH_dD^c + QH_uU^c + LH_dE^c  + LH_uN^c
                                            \nonumber \\
             {}& & \qquad   + SH_uH_d +
                     gN^cD^c + gE^cU^c + g^cU^cD^c.
\end{eqnarray}

We assume that the massless matter fields are
composed of chiral multiplets
$\Phi _i$ $(i = 1,\cdots , N_f)$ and a set
$(\delta = 1)$ of vector-like multiplets $\Phi _0$
and $\overline {\Phi }$.
In Calabi-Yau string compactification the generation
structure of matter fields is
closely linked to the topological structure
of the compactified manifold.
Here we introduce the discrete $R$-symmetry
$Z_{2k+1} \times Z_2$.
As we will see below, setting $k \geq 6$ is suitable
for explaining the mass pattern of quarks
and leptons.
The $Z_2$ symmetry is taken so as to be in accord
with the $R$-parity in the MSSM.
Therefore, hereafter the $Z_2$ symmetry is referred
to as $R$-parity.
It is assumed that the $R$-parity is conserved over
the whole energy range from
the string scale to the electroweak scale.
Supposing that ordinary quarks and leptons are
included in chiral multiplets
$\Phi _i$ $(i = 1,\cdots , N_f)$,
$R$-parity of all
$\Phi _i$ $(i = 1,\cdots , N_f)$ are set to be odd.
The assumption of $R$-parity conservation implies
that none of the chiral multiplets $\Phi _i$
$(i=1,\cdots, N_f)$ develop their
vacuum expectation values (VEVs).
Since light Higgs scalars are even under $R$-parity,
light Higgs doublets are bound to reside in
$\Phi _0$ and/or $\overline {\Phi }$.
The $D$-flatness condition and $R$-parity
conservation at large energy scales require
even $R$-parity for $\Phi _0$ and $\overline {\Phi }$
as shown in the next section.
Hence, through the spontaneous breaking of gauge
symmetry gauge superfields are possibly mixed with
the vector-like multiplets $\Phi _0$ and
$\overline {\Phi }$
but not with the chiral multiplets $\Phi _i$
$(i=1,\cdots, N_f)$.
Furthermore, no mixing occurs between the vector-like
multiplets and the chiral multiplets.

We use the $Z_{2k+1}$ symmetry as a horizontal
symmetry.
The $Z_{2k+1}$ symmetry controls
not only a large hierarchy of
the energy scales of the symmetry breaking
but also a hierarchy of effective couplings.
We denote the $Z_{2k+1}$-charges of chiral multiplets
$\Phi _i({\bf 15, 1})$ and $\Phi _i({\bf 6^*, 2})$
by $a_i$ and $b_i$ $(i = 0, 1,\cdots , N_f)$,
respectively.
In Table I, we tabulate the notations for
$Z_{2k+1}$-charges and the assignment of $R$-parity
for each superfield.
Note that the anticommuting number $\theta $ has
also a $Z_{2k+1} \times Z_2$-charge $(-1, -)$.

\vspace {5mm}
\begin{center}
\framebox [3cm] {\large \bf Table I}
\end{center}
\vspace {5mm}


\section{Spectra of vector-like multiplets}
\hspace*{\parindent}
The discrete symmetry introduced above puts
stringent constraints on both renormalizable
and nonrenormalizable interactions
in the superpotential.
To begin with, $Z_{2k+1}$-charges of
vector-like multiplets are chosen such that
both the nonrenormalizable interactions
\begin{equation}
   (S_0 {\overline S})^{2k} \quad
   ((\Phi _0{\bf (15, 1)}
       \overline {\Phi }{\bf (15^*, 1)})^{2k})
                               \nonumber
\end{equation}
and
\begin{equation}
   (N^c_0 {\overline N^c})^2 \quad
   ((\Phi _0{\bf (6^*, 2)}
       \overline {\Phi }{\bf (6, 2)})^2)
                               \nonumber
\end{equation}
are allowed
\cite{Disc}
\cite{Majorana}.
This implies that
$2k(a_0 + {\overline a}) + 2 \equiv
2(b_0 + {\overline b}) + 2 \equiv 0$ in modulus
$(2k+1)$.
Thus we set
\begin{equation}
   a_0 + {\overline a} \equiv 2, \qquad
   b_0 + {\overline b} \equiv -1  \qquad
                                {\rm mod}\ (2k+1).
\label{eqn:ab}
\end{equation}
In this case the interaction
$(S_0 {\overline S})^k (N^c_0 {\overline N^c})$
turns out to be allowed.
In contrast with this, $R$-parity conservation
forbids all of the interactions
$(S_0 {\overline S})^k (N^c_i {\overline N^c})$
$(i = 1,\cdots , N_f)$.
Incorporating the soft supersymmetry (SUSY) breaking
terms together with the $F$- and $D$-terms,
we get the scalar potential.
The scale of SUSY breaking $m_{3/2}$ is
supposed to be $O(1{\rm TeV})$.
Through the minimization of the scalar potential
we are able to detemine a ground state,
which is characterized by VEVs of
$\Phi _0$, $\overline \Phi $ and
$\Phi _i$ $(i=1,\cdots, N_f)$.
We do not address this issue here.
Instead we assume that the scalar potential
is minimized along the direction
where all chiral multiplets $\Phi _i$
$(i=1,\cdots, N_f)$ with odd $R$-parity
have vanishing VEVs.
As a result, under some conditions on soft SUSY
breaking parameters the minimization of
the scalar potential yields
\cite{Disc}
\cite{Majorana}
\begin{eqnarray}
     \langle S_0 \rangle & = &
           \langle {\overline S} \rangle \
                 \simeq \  M_S \,x, \\
     \langle N^c_0 \rangle & = &
           \langle {\overline N^c} \rangle \
                 \simeq \  M_S \,x^k,
\end{eqnarray}
up to phase factors, where
\begin{equation}
     x = \left ( \frac{m_{3/2}}{M_S} \right )
                             ^{\frac{1}{4k-2}}.
\label{eqn:x}
\end{equation}
Although for a large $k$ the parameter $x$
by itself is not a very small number,
the large hierarchy occurs by raising the number
to large powers.
Hence, $x$ becomes an efficient parameter
in describing the hierarchical structure
of the effective theory.
Further we have the inequalities
\begin{equation}
   M_S > \langle S_0 \rangle \gg
      \langle N^c_0 \rangle \gg \sqrt{m_{3/2}M_S}.
\end{equation}
Since $S_0$, ${\overline S}$, $N^c_0$ and
${\overline N^c}$ acquire VEVs along a $D$-flat
direction, $|\langle S_0 \rangle| =
|\langle {\overline S} \rangle|$
and $|\langle N^c_0 \rangle| =
|\langle {\overline N^c} \rangle|$,
SUSY is maintained down to $O(1{\rm TeV})$.
The gauge symmetry is spontaneously broken
in two steps at the scales $\langle S_0 \rangle $
and $\langle N^c_0 \rangle$ as
\begin{eqnarray}
   SU(6) \times SU(2)_R
   & \buildrel \langle S_0 \rangle \over \longrightarrow &
             SU(4)_{PS} \times SU(2)_L \times SU(2)_R \\
   & \buildrel \langle N^c_0 \rangle \over \longrightarrow &
             SU(3)_c \times SU(2)_L \times U(1)_Y,
\end{eqnarray}
where $SU(4)_{PS}$ stands for the Pati-Salam $SU(4)$
\cite{Pati}.

At the first step of the symmetry breaking
chiral superfields $Q_0$, $L_0$, ${\overline Q}$,
${\overline L}$ and $(S_0 - {\overline S})/\sqrt{2}$
are absorbed by gauge superfields.
Through the subsequent symmetry breaking
chiral superfields $U^c_0$, $E^c_0$, ${\overline U^c}$,
${\overline E^c}$ and $(N^c_0 - {\overline N^c})/\sqrt{2}$
are absorbed.
On the other hand, for components $(S_0 + {\overline S})/\sqrt{2}$
and $(N^c_0 + {\overline N^c})/\sqrt{2}$
the mass matrix is of the form
\begin{equation}
   \left(
   \begin{array}{cc}
      O(x^{4k-2})   &  O(x^{3k-1}) \\
      O(x^{3k-1})   &  O(x^{2k})
   \end{array}
   \right)
\end{equation}
in $M_S$ units.
This yields mass eigen values
\begin{equation}
   O(m_{3/2}), \ \ O(M_S x^{2k}),
\end{equation}
which correspond to the eigen states
\begin{equation}
\begin{array}{c}
   \vphantom{\bigg(}
   \frac {1}{\sqrt{2}}(S_0 + {\overline S})
       + O(x^{k-1})\frac {1}{\sqrt{2}}
                  (N^c_0 + {\overline N^c}),  \\
   \vphantom{\bigg(}
   \frac {1}{\sqrt{2}}(N^c_0 + {\overline N^c})
       + O(x^{k-1})\frac {1}{\sqrt{2}}
                  (S_0 + {\overline S}),
\end{array}
\end{equation}
respectively
\cite{Majorana}.
The discrete symmetry $Z_{2k+1}$ is
broken together with $SU(6) \times SU(2)_R$
by $\langle S_0 \rangle $,
while there remains an unbroken $Z_2$-symmetry
referred to $R$-parity conservation.

In order to stabilize the weak-scale hierarchy
we put an additional requirement that
the interaction
\begin{equation}
   (S_0 {\overline S})^{2k-1} S_0 H_{u0} H_{d0}
\end{equation}
is allowed in the superpotential.
This condition translates into
\begin{equation}
     2b_0 \equiv {\overline a} \ \ \ {\rm mod}\ (2k+1).
\label{eqn:abb}
\end{equation}
 From Eqs.(\ref{eqn:ab}) and (\ref{eqn:abb})
the trilinear coupling
${\overline S}\,{\overline H_u}{\overline H_d}$
becomes to be allowed.
Therefore, the superpotential of Higgs doublet
in vector-like multiplets has the form
\begin{equation}
    W_H \sim {\overline S}\,{\overline H_u}{\overline H_d}
              + (S_0 {\overline S})^k
                           (H_{u0} {\overline H_u}
                 + H_{d0} {\overline H_d}) \\
              + (S_0 {\overline S})^{2k-1}
                              S_0 H_{u0} H_{d0}.
\end{equation}
When $S_0$ and $\overline S$ develop the non-zero
VEVs,
the superpotential induces the mass matrix
of $H_{u0}$, $H_{d0}$,
${\overline H_u}$ and ${\overline H_d}$
\begin{equation}
\begin{array}{r@{}l}
    \vphantom{\bigg(}  &  \begin{array}{ccc}
        \quad {\overline H_u} \quad & \quad H_{d0} &
       \end{array} \\
    \begin{array}{l}
      {\overline H_d}  \\  H_{u0} \\
    \end{array}
    &
 \left(
   \begin{array}{cc}
      O(x)        &  O(x^{2k}) \\
      O(x^{2k})   &  O(x^{4k-1})
   \end{array}
 \right)
\end{array}
\end{equation}
in $M_S$ units,
which leads to the mass eigen values
\begin{equation}
   O(M_S \,x), \ \ O(M_S \,x^{4k-1}) = O(m_{3/2} \,x).
\end{equation}
Namely we have the $\mu $-term with $\mu =
O(m_{3/2} \,x) = O(10^{2 \sim 3}{\rm GeV})$
\cite{Aligned}\cite{Tadpole}.
Light Higgs states are given by
\begin{equation}
     H_{u0} + O(x^{2k-1}){\overline H_d}, \qquad
     H_{d0} + O(x^{2k-1}){\overline H_u}.
\end{equation}
The components of $\overline{H_d}$ and
$\overline{H_u}$ in light Higgses are extremely small
because $x^{2k-1} = (m_{3/2}/M_S)^{1/2} = O(10^{-8})$.
Note that the product $H_{u0} H_{d0}$ has a nonzero
$Z_{2k+1}$-charge.
In contrast with the present model,
in a solution of the $\mu $-problem proposed in
Ref.\cite{Miu}
the $R$-charge of the product of light Higgses
has to be zero.

The remaining components in $\Phi _0$ and
$\overline {\Phi }$, i.e. $g_0$, $g^c_0$, $D^c_0$ and
$\overline g$, ${\overline g^c}$, ${\overline D^c}$
are down-type color-triplet fields.
In the present model the spectra of color-triplet
Higgses are quite different from those of
doublet Higgses.
In general down-type colored fields
potentially mediate proton decay.
However, since chiral multiplets
$\Phi _i$ $(i=1,\cdots , N_f)$
which contain ordinary quarks and leptons have
odd $R$-parity,
lepto-quark chiral superfields mediating proton decay
should be even under $R$-parity.
Such superfields are limited to $g_0$, $g^c_0$
and $D^c_0$ in $\Phi _0$.
If we put
\begin{equation}
   3a_0 \equiv -2 - 2q \qquad {\rm mod}\ (2k+1)
\label{eqn:a0}
\end{equation}
with $0 \leq q \lsim k$,
then the interaction
$(S_0 {\overline S})^q S_0 g_0 g^c_0$ is allowed.
 From Eqs.(\ref{eqn:ab}),
(\ref{eqn:abb}) and (\ref{eqn:a0})
we can summarize as
\begin{equation}
\begin{array} {rcl}
    3a_0 & \equiv & -2 - 2q, \qquad \ \
    {\overline a} \equiv  2 - a_0,  \\
    b_0 & \equiv & 2 + a_0 + q, \qquad
    {\overline b} \equiv  -3 - a_0 - q  \qquad
              {\rm mod}\ (2k+1).
\end{array}
\label{eqn:aabb}
\end{equation}
Hence, under the discrete $R$-symmetry
the superpotential of down-type colored fields is
\begin{eqnarray}
     W_g & \sim & (S_0 \overline S)^q S_0 g_0 g^c_0 +
                (S_0 \overline S)^{2k-4-q}
             (\overline S \overline g \,{\overline g^c}
             + N^c_0 \overline S D^c_0 {\overline g^c})
                                     \nonumber  \\
       & &   + (S_0 \overline S)^{2k-1} (g_0 \overline g
             +  g^c_0 {\overline g^c} + g_0 N^c_0 D^c_0 )
             + \overline g {\overline N^c} {\overline D^c}
                                     \nonumber  \\
       & &   + (S_0 \overline S)^k D^c_0 {\overline D^c}
             + (S_0 \overline S)^{q+2} S_0 {\overline N^c}
                g^c_0 {\overline D^c}.
\end{eqnarray}
When $S_0$, $\overline S$, $N^c_0$ and ${\overline N^c}$
develop non-zero VEVs,
the superpotential induces the mass matrix
\begin{equation}
\begin{array}{r@{}l}
   \vphantom{\bigg(}    &  \begin{array}{cccc}
            \qquad \ g^c_0 \qquad \ & \qquad \ \overline{g}
                         \qquad \ & \qquad \ D^c_0  &
         \end{array}  \\
     \begin{array}{l}
        g_0  \\  {\overline g^c}  \\  {\overline D^c} \\
     \end{array}
      &
\left(
  \begin{array}{ccc}
  O(x^{2q+1})   &  O(x^{4k-2})    &  O(x^{5k-2})    \\
  O(x^{4k-2})   &  O(x^{4k-7-2q}) &  O(x^{5k-7-2q}) \\
  O(x^{k+5+2q}) &  O(x^k)         &  O(x^{2k})
  \end{array}
\right)
\end{array}
\end{equation}
in $M_S$ units.
This leads to mass eigen values
\begin{equation}
    O(M_S \,x^{2q+1}), \qquad O(M_S \,x^k),
                \qquad O(m_{3/2} \,x^{k-5-2q}).
\label{eqn:g0}
\end{equation}
The eigen state associated with
a mass of $O(M_S \,x^{2q+1})$
consists mainly of $g_0$ and $g^c_0$.
On the other hand, the lightest states consists
mainly of ${\overline g^c}$ and $D^c_0$.
When $k-5 > 2q$, the lightest eigen value is
smaller than $m_{3/2}$.
Since there is no experimental evidence for extra
colored particles,
we take $k-5 < 2q \lsim 2k$.
It is worth noting that these down-type colored
fermions are $SU(2)_L$-singlet and odd under $R$-parity.
Due to $R$-parity conservation this particle should
be produced in pair or together with a superparticle.
A question arises as to whether or not
these colored particles bring about
unacceptably large nucleon-decay rates.
Although the detailed study of this issue
will be presented in another paper,
we add a few words here.
If we assign appropriate
discrete-charges to chiral multiplets $\Phi _i$,
both the dimension-five and -six operators
mediating proton decay
are suppressed enough to avoid fast proton decay.


\section{Quark masses and the CKM matrix}
\hspace*{\parindent}
Next we turn to mass matrices for chiral multiplets
$\Phi _i$ $(i=1, \cdots , N_f)$.
Due to $R$-parity conservation
$\Phi _i$ $(i= 1, \cdots , N_f)$
are not mixed with vector-like multiplets
$\Phi _0$ and $\overline {\Phi }$.
The superpotential of up-type quarks is given by
\begin{equation}
   W_U \sim (S_0 \overline {S})^{m_{ij}} Q_i U^c_j H_{u0}
          \qquad (i, j = 1,\cdots , N_f),
\end{equation}
where the exponents $m_{ij}$ are integers
in the range $0 \leq m_{ij} < 2k+1$.
Recall that light Higgs doublets are almost
$H_{u0}$ and $H_{d0}$.
Under the $Z_{2k+1}$-symmetry the exponent $m_{ij}$
is determined by the condition
\begin{equation}
   2 m_{ij} + b_0 + a_i + b_j + 2 \equiv 0
                        \qquad {\rm mod}\ (2k+1).
\end{equation}
Instead of $a_i$ and $b_i$
$(i = 1,\cdots , N_f)$,
if we use new parameters $c_i$ and $d_i$
defined by
\begin{equation}
    c_i \equiv a_0 - a_i, \qquad
    d_i \equiv a_0 - b_i,
\end{equation}
then the above condition is rewritten as
\begin{equation}
   2 m_{ij} \equiv q - 2 + c_i + d_j
                        \qquad {\rm mod}\ (2k+1).
\label{eqn:mij}
\end{equation}
The mass matrix of up-type
quarks is described by an $N_f \times N_f$ matrix
$M$ with elements
\begin{equation}
         M_{ij} = O(x^{2 m_{ij}})
\end{equation}
multiplied by $v_u = \langle H_{u0} \rangle $.
 From Eq.(\ref{eqn:mij})
the matrix $M$ is generally asymmetric.
When we adopt appropriate unitary matrices
${\cal V}_u$ and ${\cal U}_u$,
the matrix
\begin{equation}
    {\cal V}_u^{-1} M \,{\cal U}_u
\end{equation}
becomes diagonal.

Under $SU(6) \times SU(2)_R$ gauge symmetry
down-type quarks and leptons share
the nonrenormalizable terms in common with
up-type quarks.
Namely we get
\begin{equation}
   W \sim (S_0 \overline {S})^{m_{ij}}
       \{ Q_i D^c_j H_{d0} + L_i N^c_j H_{u0}
           + L_i E^c_j H_{d0} \}.
\end{equation}
For down-type quarks, however, the mixings between
$g^c$ and $D^c$ should be taken into account
at energies below the scale $\langle N_0^c \rangle$.
In fact, the superpotential of down-type colored fields
is of the form
\begin{equation}
   W_D \sim (S_0 \overline {S})^{z_{ij}} S_0 g_i g^c_j
           + (S_0 \overline {S})^{m_{ij}}
       ( N^c_0 g_i + H_{d0} Q_i) D^c_j,
\end{equation}
where the exponents $z_{ij}$ are determined by
\begin{equation}
    2 z_{ij} \equiv 2q + c_i + c_j \qquad
                           {\rm mod}\ (2k+1)
\end{equation}
in the range $0 \leq z_{ij} < 2k+1$.
When the gauge group $G$ is smaller than
$SU(6) \times SU(2)_R$,
up-type quarks, down-type quarks and leptons
do not always share the nonrenormalizable terms
in common.
The present choice of $G$ makes the model
transparent.
In terms of an $N_f \times N_f$ matrix $Z$
with elements
\begin{equation}
     Z_{ij} = O(x^{2z_{ij}}),
\end{equation}
a mass matrix of down-type colored fields is
written as
\begin{equation}
\begin{array}{r@{}l}
   \vphantom{\bigg(}   &  \begin{array}{ccc}
          \quad   g^c   &  \quad D^c  &
        \end{array}  \\
\widehat{M}_d =
   \begin{array}{l}
        g  \\  D  \\
   \end{array}
     &
\left(
  \begin{array}{cc}
      x Z   &    x^k M    \\
       0    &  \rho _d M
  \end{array}
\right)
\end{array}
\label{eqn:Mdh}
\end{equation}
in $M_S$ units below the scale
$\langle N^c_0 \rangle $,
where $\rho _d = \langle H_{d0} \rangle /M_S = v_d /M_S$.
This $\widehat{M}_d$ is a $2N_f \times 2N_f$ matrix
and can be diagonalized by a bi-unitary transformation
as
\begin{equation}
    \widehat{\cal V}_d^{-1} \widehat{M}_d \,
                        \widehat{\cal U}_d.
\label{eqn:Md}
\end{equation}
An explicit form of $\widehat{M}_d$ shows
mixings between $g^c$ and $D^c$.
This type of mixings does not occur for up-type
quarks.
 From Eqs. (\ref{eqn:Mdh}) and (\ref{eqn:Md})
the matrix
\begin{equation}
    \widehat{\cal V}_d^{-1} \widehat{M}_d
         \widehat{M}_d^{\dag } \widehat{\cal V}_d =
    \widehat{\cal V}_d^{-1}
      \left(
      \begin{array}{cc}
          A_d + B_d    &  \epsilon B_d   \\
        \epsilon B_d   &  \epsilon ^2 B_d
      \end{array}
      \right)
    \widehat{\cal V}_d,
\end{equation}
is diagonal, where
\begin{equation}
  A_d = x^2 Z Z^{\dag}, \qquad B_d = x^{2k} M M^{\dag},
                    \qquad \epsilon = \rho _d \,x^{-k}.
\end{equation}

In view of the smallness of the parameter
$\epsilon = O(10^{-12})$,
we use the perturbative method
in solving the eigenvalue problem.
It follows that the eigen equation is approximately
separated into two pieces.
For heavy states the eigen equation becomes
\begin{equation}
   \det \left( A_d + B_d - \frac {\eta }{M_S^2}
                \right) = 0.
\end{equation}
Solving this equation of a variable $\eta $,
we obtain masses squared for $N_f$ heavy states.
The other $N_f$ states are light and their masses
squared are given by solving the eigen equation
\begin{equation}
   \det \left( x^{-2k}( A_d^{-1} + B_d^{-1} )^{-1}
         - \frac {\eta }{v_d^2} \right) = 0.
\end{equation}
This equation is derived in $\epsilon ^2$ order of
the perturbative expansion.
The light states correspond to observed down-type
quarks.
If the mixing between $g^c$ and $D^c$ is sizable,
mass pattern of down-type quarks is possibly
changed from that of up-type quarks.
Thus in our model, property (ii) of observed
fermion masses is attributable to this mixing
mechanism.

The $2N_f \times 2N_f$ unitary matrices
$\widehat{\cal{V}}_d$ and $\widehat{\cal{U}}_d$
are
\begin{eqnarray}
   \widehat{\cal V}_d & \simeq & \left(
   \begin{array}{cc}
      {\cal W}_d   &  -\epsilon (A_d + B_d)^{-1}
                                    B_d {\cal V}_d \\
      \epsilon B_d (A_d + B_d)^{-1} {\cal W}_d    &
                                          {\cal V}_d
   \end{array}
                       \right), \\
   \widehat{\cal U}_d & \simeq & \left(
   \begin{array}{cc}
      xZ^{\dag} {\cal W}_d \,(\Lambda _d^{(0)})^{-1/2}
         &  -(xZ)^{-1} {\cal V}_d \,
                               (\Lambda _d^{(2)})^{1/2}  \\
      x^k M^{\dag} {\cal W}_d \,(\Lambda _d^{(0)})^{-1/2}
         &  (x^kM)^{-1} {\cal V}_d \,
                                 (\Lambda _d^{(2)})^{1/2}
   \end{array}
   \right),
\end{eqnarray}
respectively.
Here ${\cal W}_d$ and ${\cal V}_d$ are
$N_f \times N_f$ unitary matrices
which are determined such that the matrices
\begin{equation}
    {\cal W}_d^{-1}(A_d + B_d){\cal W}_d
                          = \Lambda _d^{(0)}, \qquad
    {\cal V}_d^{-1}(A_d^{-1} + B_d^{-1})^{-1}
              {\cal V}_d = \Lambda _d^{(2)}
\end{equation}
become diagonal.
As a consequence we have a nontrivial CKM matrix
\begin{equation}
   V^{CKM} = {\cal V}_u^{-1} \,{\cal V}_d.
\end{equation}
Note that ${\cal V}_u$ is determined such that
${\cal V}_u^{-1} B_d {\cal V}_u$ is diagonal.
In the case when the relation
\begin{equation}
    | \det (A_d+B_d)| \simeq | \det A_d |
                             \gg | \det B_d |
\end{equation}
is satisfied,
the mixing is small and we have
\begin{equation}
    ( A_d^{-1} + B_d^{-1} )^{-1} \simeq B_d.
\end{equation}
This implies that mass pattern of down-type
quarks is the same as that of up-type quarks
and that ${\cal V}_d \simeq {\cal V}_u$.
Therefore, in this case $V^{CKM}$ becomes
almost a unity matrix.
To get a phenomenologically viable solution,
large mixings between $g^c$ and $D^c$
are preferable.
A concrete example is exhibited in section 6.


\section{Spectra of leptons}
\hspace*{\parindent}
Let us now study the mass matrices for lepton sector,
in which $L$-$H_d$ mixing occurs at energies
below the scale $\langle N_0^c \rangle$.
As mentioned in section 2,
$H_{ui}$ and $H_{di}$ $(i=1,\cdots, N_f)$
in chiral multiplets do not
develop their VEVs.
It follows that there exist no mixings among
$SU(2)_L \times U(1)_Y$ gauge superfields,
$H_{ui}$ and $H_{di}$ $(i=1,\cdots, N_f)$.
Since both $L$ and $H_d$ are $SU(2)_L$-doublets,
the CKM matrix for lepton sector becomes a unit matrix
irrespective of the magnitude of $L$-$H_d$ mixing.
For charged leptons the superpotential is
\begin{equation}
   W_E \sim (S_0 \overline {S})^{h_{ij}} S_0 H_{di} H_{uj}
           + (S_0 \overline {S})^{m_{ij}}
             L_i ( N^c_0 H_{uj} + H_{d0} E^c_j),
\end{equation}
where the exponents $h_{ij}$ are integers in the range
$0 \leq h_{ij} < 2k+1$ and satisfy
\begin{equation}
    2 h_{ij} \equiv 2q + d_i + d_j \qquad
                           {\rm mod}\ (2k+1).
\end{equation}
As before, we introduce an $N_f \times N_f$ matrix
$H$ with elements
\begin{equation}
    H_{ij} = O(x^{2h_{ij}}).
\end{equation}
The mass matrix for charged leptons
has the form
\begin{equation}
\begin{array}{r@{}l}
   \vphantom{\bigg(}   &  \begin{array}{ccc}
          \quad   H_u^+   &  \quad E^{c+}  &
        \end{array}  \\
\widehat{M}_l =
   \begin{array}{l}
        H_d^-  \\  L^-  \\
   \end{array}
     &
\left(
  \begin{array}{cc}
      x H    &    0       \\
      x^k M  &  \rho _d M
  \end{array}
\right)
\end{array}
\label{eqn:Mlh}
\end{equation}
in $M_S$ units.
This $\widehat{M}_l$ is also a $2N_f \times 2N_f$
matrix and can be diagonalized by
a bi-unitary transformation as
\begin{equation}
    \widehat{\cal V}_l^{-1} \widehat{M}_l \,
                        \widehat{\cal U}_l.
\label{eqn:Ml}
\end{equation}
 From Eqs.(\ref{eqn:Mlh}) and (\ref{eqn:Ml})
the matrix
\begin{equation}
    \widehat{\cal U}_l^{-1} \widehat{M}_l^{\dag }
              \widehat{M}_l \, \widehat{\cal U}_l =
    \widehat{\cal U}_l^{-1}
      \left(
      \begin{array}{cc}
          A_l + B_l    &  \epsilon B_l   \\
        \epsilon B_l   &  \epsilon ^2 B_l
      \end{array}
      \right)
    \widehat{\cal U}_l
\end{equation}
is diagonal, where
\begin{equation}
  A_l = x^2 H^{\dag} H, \qquad B_l = x^{2k} M^{\dag} M.
\end{equation}

The analysis is parallel to that of down-type
quark masses in the previous section.
We have the eigen equation
\begin{equation}
   \det \left( A_l + B_l - \frac {\eta }{M_S^2}
                \right) = 0
\label{eqn:ABl}
\end{equation}
for heavy states.
For $N_f$ light states their masses
squared are given by the eigen equation
\begin{equation}
   \det \left( x^{-2k}( A_l^{-1} + B_l^{-1} )^{-1}
         - \frac {\eta }{v_d^2} \right) = 0.
\label{eqn:AB-l}
\end{equation}
The light states correspond to observed
charged leptons.
Due to $L$-$H_d$ mixings
mass pattern of charged leptons could be
changed from that of up-type quarks.
Introducing appropriate unitary matrices
${\cal W}_l$ and ${\cal V}_l$,
we can diagonalize $(A_l + B_l)$ and
$(A_l^{-1} + B_l^{-1})^{-1}$ as
\begin{equation}
  {\cal W}_l^{-1} (A_l + B_l) {\cal W}_l
                     = \Lambda _l^{(0)}, \qquad
  {\cal V}_l^{-1} (A_l^{-1} + B_l^{-1})^{-1}
                 {\cal V}_l = \Lambda _l^{(2)},
\end{equation}
where $\Lambda _l^{(0)}$ and $\Lambda _l^{(2)}$
are diagonal $N_f \times N_f$ matrices.
Masses of charged leptons are written as
\begin{equation}
   m_{li}^2 = v_d^2 \left( x^{-2k}
            \Lambda _l^{(2)} \right)_{ii}
                   \qquad (i = 1,\cdots, N_f).
\end{equation}
Explicit forms of ${\widehat{\cal V}}_l$
and ${\widehat{\cal U}}_l$ are
\begin{eqnarray}
   \widehat{\cal V}_l & \simeq & \left(
   \begin{array}{cc}
      xH {\cal W}_l \,(\Lambda _l^{(0)})^{-1/2}
         &  -(xH^{\dag})^{-1} {\cal V}_l
                     \,(\Lambda _l^{(2)})^{1/2}  \\
      x^k M {\cal W}_l \,(\Lambda _l^{(0)})^{-1/2}
         &  (x^kM^{\dag})^{-1} {\cal V}_l
                         \,(\Lambda _l^{(2)})^{1/2} \\
   \end{array}
                       \right), \\
   \widehat{\cal U}_l & \simeq & \left(
   \begin{array}{cc}
      {\cal W}_l   &  -\epsilon (A_l + B_l)^{-1}
                                     B_l {\cal V}_l \\
      \epsilon B_l (A_l + B_l)^{-1} {\cal W}_l    &
                                           {\cal V}_l
   \end{array}
                       \right).
\end{eqnarray}

We now proceed to study mass pattern of
neutral leptons.
For neutral fields the superpotential is
of the form
\begin{eqnarray}
   W_N & \sim & (S_0 \overline {S})^{h_{ij}}
                              S_0 H_{di} H_{uj}
           + (S_0 \overline S)^{m_{ij}}
             L_i ( N^c_0 H_{uj} + H_{u0} N^c_j)
                                        \nonumber  \\
       & & \quad + (S_0 \overline S)^{s_{ij}}
                (S_i \overline S) (S_j \overline S)
             + (S_0 \overline S)^{t_{ij}}
                (S_i \overline S)
                 (N^c_j {\overline N^c})  \nonumber \\
       & & \quad + (S_0 \overline S)^{n_{ij}}
                       (N^c_i {\overline N^c})
                          (N^c_j {\overline N^c}),
\end{eqnarray}
where the exponents $s_{ij}$, $t_{ij}$ and $n_{ij}$
are determined by
\begin{eqnarray}
  2s_{ij} & \equiv & 2z_{ij} - 2q - 6,
                                  \nonumber \\
  2t_{ij} & \equiv & 2m_{ij} + 1,
                       \qquad {\rm mod}\ (2k+1)  \\
  2n_{ij} & \equiv & 2h_{ij} + 4
                                  \nonumber
\end{eqnarray}
in the range $0 \leq s_{ij}, t_{ij}, n_{ij} < 2k+1$.
Introducing $N_f \times N_f$ matrices $S$, $T$ and $N$
with elements
\begin{equation}
     S_{ij} = O(x^{2s_{ij}}), \quad
     T_{ij} = O(x^{2t_{ij}}), \quad
     N_{ij} = O(x^{2n_{ij}}),
\end{equation}
we have a $5N_f \times 5N_f$ mass matrix
\begin{equation}
\begin{array}{r@{}l}
   \vphantom{\bigg(}   &  \begin{array}{cccccc}
          \quad \, H_u^0   &  \ \  H_d^0  &  \quad L^0
                          &  \quad N^c   & \quad \quad  S  &
        \end{array}  \\
\widehat{M}_N =
   \begin{array}{l}
        H_u^0  \\  H_d^0  \\  L^0  \\  N^c  \\  S  \\
   \end{array}
     &
\left(
  \begin{array}{ccccc}
      0     &  x H  &  x^k M^T   &     0     &    0   \\
      x H   &   0   &    0       &     0     &    0   \\
      x^k M &   0   &    0       & \rho _u M &    0   \\
      0     &   0   & \rho _u M^T
                            & x^{2k} N  & x^{k+1} T^T \\
      0     &   0   &    0       & x^{k+1} T & x^2 S  \\
  \end{array}
\right)
\end{array}
\label{eqn:Mn}
\end{equation}
in $M_S$ units for neutral sector.
Since $SU(2)_L$ symmetry is preserved above the
electroweak scale,
the eigen equation for $2N_f$ heavy states is
the same as Eq.(\ref{eqn:ABl}).
For another $2N_f$ states which are
$G_{st}$-neutral
we have the eigen equation
\begin{equation}
   \det \left( \widehat{M}_{NS} -
               \frac {\eta }{M_S^2} \right) = 0,
\end{equation}
where $\widehat{M}_{NS}$ is a submatrix of
$\widehat{M}_N$ defined by
\begin{equation}
   \widehat{M}_{NS} = \left(
   \begin{array}{cc}
       x^{2k} N  &  x^{k+1} T^T \\
       x^{k+1} T &  x^2 S
   \end{array}
   \right).
\end{equation}
Right-handed neutrinos with Majorana masses are
contained in the latter $2N_f$ states.
For light neutrinos there appears the same Dirac
mass matrix as that for charged leptons.
When we have large Majorana masses of
right-handed neutrinos,
through seesaw mechanism
\cite{Seesaw}
light neutrino masses are given by
\begin{equation}
   m_{\nu i} = \frac {m_{li}^2}{M_S \,x^{2k}}
             \left( \frac {v_u}{v_d} \right)^2
             \left( {\cal V}_l^{-1} \Delta _N
                        {\cal V}_l \right)_{ii}
\label{eqn:Majo}
\end{equation}
with
\begin{equation}
   \Delta _N = (N-T^T S^{-1} T )^{-1}.
\end{equation}


\section{A two-generation toy model}
\hspace*{\parindent}
In this section we take up a two-generation
case $(N_f = 2)$ as a simple example to
illustrate general features
of the mixing mechanism.
A two-generation toy model will provide a guide
to constructing a phenomenologically viable
model.
We choose $Z_{23} \times Z_2$ $(k=11)$ as
a discrete $R$-symmetry.
By substituting $k=11$, $m_{3/2} = 1$TeV and
$M_S = 10^{18}$GeV for Eq.(\ref{eqn:x}),
the parameter $x$ becomes
\begin{equation}
    x \simeq 10^{-0.36} \simeq \frac {1}{2.3}.
\end{equation}
Hence, the energy scales of
symmetry breaking are
\begin{eqnarray}
     \langle S_0 \rangle & \simeq
                          & 10^{17.6}{\rm GeV}, \\
     \langle N^c_0 \rangle & \simeq
                              & 10^{14.0}{\rm GeV}.
\end{eqnarray}
In the case $q=9$ the $Z_{23}$-charges of
vector-like multiplets are settled from
Eq.(\ref{eqn:aabb}) as
\begin{equation}
  a_0 \equiv 1, \quad
               {\overline a} \equiv 1, \quad
  b_0 \equiv 12, \quad
               {\overline b} \equiv 10
                         \qquad {\rm mod}\ 23.
\end{equation}

For chiral multiplets $\Phi _1$ and $\Phi _2$,
for instance, we choose
\begin{equation}
  c_1 \equiv 4, \quad c_2 \equiv 2, \quad
  d_1 \equiv 5, \quad d_2 \equiv 22,
\end{equation}
i.e.,
\begin{equation}
  a_1 \equiv 20,  \quad a_2 \equiv 22, \quad
  b_1 \equiv 19,  \quad b_2 \equiv 2
\end{equation}
in modulus 23.
Under this assignment the matrices $Z$, $M$ and $H$
are given by
\begin{eqnarray}
   Z & = & \lambda ^{11} \left(
      \begin{array}{cc}
          O(\lambda ^2)  &  O(\lambda )  \\
          O(\lambda )    &  O(1)         \\
      \end{array}
      \right), \\
   M & = & \lambda ^4 \left(
      \begin{array}{cc}
          O(\lambda ^4)  &  O(\lambda )  \\
          O(\lambda ^3)  &  O(1)         \\
      \end{array}
      \right), \\
   H & = & \lambda ^8 \left(
      \begin{array}{cc}
          O(\lambda ^6)  &  O(\lambda ^3)  \\
          O(\lambda ^3)  &  O(1)
      \end{array}
      \right),
\end{eqnarray}
where
\begin{equation}
    \lambda = x^2 \sim \frac {1}{5}.
\end{equation}
The unitary matrix ${\cal V}_u$
which diagonalizes $M M^{\dag } = x^{-2k}B_d$,
is of the form
\begin{equation}
    {\cal V}_u =
      \left(
      \begin{array}{cc}
        1 - O(\lambda ^2)  &  O(\lambda )    \\
        O(\lambda )        &  1 - O(\lambda ^2)
      \end{array}
      \right).
\end{equation}
As a result we obtain
\begin{equation}
      {\cal V}_u^{-1} M M^{\dag } {\cal V}_u =
      \left(
      \begin{array}{cc}
        O(\lambda ^{16})  &   0    \\
        0                 &   O(\lambda ^8)
      \end{array}
      \right).
\end{equation}
This means that masses of up-type quarks are
\begin{equation}
  m_{u1} = v_u \,O(\lambda ^8), \quad
  m_{u2} = v_u \,O(\lambda ^4),
\end{equation}
which correspond to $u$-quark and $c$-quark,
respectively.
As mentioned in section 4,
masses of down-type quarks are determined by
the eigen values of
$x^{-2k} (A_d^{-1} + B_d^{-1})^{-1}$.
Incidentally, we have a relation
\begin{equation}
     (A^{-1} + B^{-1})^{-1} =
           \frac {1}{\det (A+B)} \left(
            A \,(\det B) + B \,(\det A) \right)
\end{equation}
for non-singular $2 \times 2$ matrices $A$ and $B$.
It is efficient to apply this relation to
the present case.
A simple calculation yields
\begin{eqnarray}
   {} & & \det (x^{-2k} A_d)
                =  x^{4-4k} |\det Z|^2
                            = O(\lambda ^{28}), \\
   {} & & \det (x^{-2k} B_d)  =  |\det M|^2
                            = O(\lambda ^{24}), \\
   {} & & \det (x^{-2k} (A_d + B_d))
                           =  O(\lambda ^{22}).
\end{eqnarray}
It follows that
\begin{eqnarray}
   x^{-2k} (A_d^{-1} + B_d^{-1})^{-1} & \simeq &
               O(\lambda ^2) x^{-2k} A_d
                   + O(\lambda ^6) x^{-2k} B_d  \\
     & = & \left(
     \begin{array}{cc}
       O(\lambda ^{16}) & O(\lambda ^{15}) \\
       O(\lambda ^{15}) & O(\lambda ^{14})
     \end{array}
     \right).
\end{eqnarray}
In addition we get
\begin{equation}
     \det \left(
         x^{-2k} (A_d^{-1} + B_d^{-1})^{-1}
           \right) = O(\lambda ^{30}).
\end{equation}
This implies that two eigen values of
$x^{-2k} (A_d^{-1} + B_d^{-1})^{-1}$
are $O(\lambda ^{16})$ and $O(\lambda ^{14})$.
Thus masses of down-type quarks are
\begin{equation}
   m_{d1} = v_d \,O(\lambda ^8), \quad
   m_{d2} = v_d \,O(\lambda ^7),
\end{equation}
which correspond to $d$-quark and $s$-quark,
respectively.
It is worth emphasizing that mass pattern of
down-type quarks is changed from that of
up-type quarks through $g^c$-$D^c$ mixing mechanism.
Further the unitary matrix ${\cal V}_d$ which
diagonalizes $(A_d^{-1} + B_d^{-1})^{-1}$,
is expressed as
\begin{equation}
    {\cal V}_d =
      \left(
      \begin{array}{cc}
        1  - O(\lambda ^2)  &  O(\lambda )    \\
        O(\lambda )         &  1  - O(\lambda ^2)
      \end{array}
      \right).
\end{equation}
Although corresponding elements of the matrices
${\cal V}_u$ and ${\cal V}_d$
are in the same order of $\lambda $,
the coefficients of $\lambda $ in off-diagonal
elements of ${\cal V}_u$ and ${\cal V}_d$ are
generally different from each other.
Hence we obtain a nontrivial CKM matrix
\begin{equation}
    V^{CKM} = {\cal V}_u^{-1} {\cal V}_d =
      \left(
      \begin{array}{cc}
        1 - O(\lambda ^2)  &  O(\lambda )    \\
        O(\lambda )        &  1 - O(\lambda ^2)
      \end{array}
      \right).
\end{equation}
Consequently, the parameter
$\lambda = x^2 \sim 1/5$
corresponds to $\sin \theta _C$,
where $\theta _C$ is the Cabibbo angle.

For lepton sector a similar
calculation is carried out.
Under the present assignment of $Z_{23}$-charges
we get
\begin{equation}
    (A_l^{-1} + B_l^{-1})^{-1} \simeq
       O(\lambda ^4) A_l + O(\lambda ^4) B_l.
\end{equation}
It follows that masses of charged leptons are
\begin{equation}
   m_{l1} = v_d \,O(\lambda ^9), \qquad
   m_{l2} = v_d \,O(\lambda ^5).
\end{equation}
In this example of two-generation cases
neutrino Majorana masses are too small.


\section{Summary}
\hspace*{\parindent}
In the context of Calabi-Yau string model
with Kac-Moody level-one we have shown
a possibility that characteristic pattern of
quark-lepton masses and the CKM matrix have their
origin in the discrete $R$-symmetry and mixing
mechanism.
In this paper we have chosen $Z_{2k+1} \times Z_2$
symmetry as an example of the discrete $R$-symmetry.
The $Z_2$-symmetry is assumed so as to be in
accord with the $R$-parity in the MSSM
and to be unbroken at scales larger than
the electroweak scale.
The vector-like multiplets $\Phi _0$,
$\overline \Phi $ and the chiral multiplets
$\Phi _i$ $(i=1,\cdots, N_f)$ are assigned to
even and odd $R$-parity, respectively.
Under this assignment no mixing occurs between
the vector-like multiplets and the chiral multiplets.
The $R$-parity conservation plays an important role
in constructing a phenomenologically viable model.
The $Z_{2k+1}$ symmetry is used as a horizontal
symmetry.
The $Z_{2k+1}$ symmetry controls a large hierarchy
of the energy scales of the symmetry breaking
and particle spectra.
The assignment of $Z_{2k+1}$-charges to chiral
multiplets is of great importance in explaining
the observed hierarchical pattern of
quark-lepton masses.
The mass hierarchy of up-type quarks is a direct
result of the horizontal discrete symmetry.
On the other hand, for down-type quarks there appears
a mixing between $D^c$ and $g^c$.
Due to this mixing mass pattern of down-type quarks
is possibly different from that of up-type quarks.
Further we obtain a nontrivial CKM matrix.
For lepton sector $L$-$H_d$ mixing occurs.
Hence, mass pattern of charged leptons also could
be changed from that of up-type quarks.
However, the CKM matrix for lepton sector amounts to a
unit matrix irrespectively of the magnitude of
$L$-$H_d$ mixing.
This is because both $L$ and $H_d$ are
$SU(2)_L$-doublets.
Further, seesaw mechanism could be at work
for neutrinos.

To illustrate general features of the model
we have exhibited an example in two-generation
case.
In order to render the model phenemenologically
viable we need to study three-generation cases.
In this paper we have presented the basic idea
of explaining hierarchical pattern of quark-lepton
masses and the CKM matrix.
A search for a viable model with three
generations will be done in another paper.
Furthermore, several problems remain to be solved.
As mentioned above, we have assumed
the $R$-parity conservation.
However, a question arises as to
why the $Z_2$-symmetry associated with
the $R$-parity remains unbroken
down to the electroweak scale.
To this end, we need to study the minimization
of the scalar potential in detail.
So as to be consistent with the proton stability,
it is required that lepto-quark chiral
superfields are heavy and that
products of effective Yukawa couplings
which take part in the proton decay
are sufficiently small.
Through the detailed study of three-generation
cases we anticipate finding solutions
in which nucleon-decay rates are acceptably
suppressed.
In addition, in this paper we have chosen
$SU(6) \times SU(2)_R$ as a gauge group
to be specific.
In the light of the choice of $G$
we have to explore whether or not
the unification of gauge couplings occurs
around the string scale.
Unsolved problems associated with these subjects
will be studied elsewhere.


\newpage

\begin{center}
{\large {\bf Table I  \\ }}
\vspace {1cm}
\begin{tabular}{|c|cc|cccc|} \hline
\phantom{\Bigg(} & $\overline{\Phi }$ &  $\Phi _0$
       &  $\Phi _1$  &  $\Phi _2$  & $\cdots$
             & $\Phi _{N_f}$  \\
                                              \hline
\phantom{\Bigg(} $({\bf 15, 1})$ \quad &  $(\overline{a}, +)$
             &  $(a_0, +)$
                      &  $(a_1, -)$  &  $(a_2, -)$
                         &  $\cdots$ & $(a_{N_f}, -)$  \\
\phantom{\Bigg(} $({\bf 6^*, 2})$ \quad &  $(\overline{b}, +)$
             &  $(b_0, +)$
                      &  $(b_1, -)$  &  $(b_2, -)$
                         &  $\cdots$ & $(b_{N_f}, -)$  \\
                                              \hline
\end{tabular}
\end{center}

\vspace{3cm}

{\large Table Caption} \\
\vspace{3mm}

{\bf Table I} \qquad
The numbers $a_i$ and $b_i$ $(i = 0, 1, \cdots, N_f)$
in the parentheses represent the $Z_{2k+1}$-charges
of chiral superfields $\Phi ({\bf 15, 1})$ and
$\Phi ({\bf 6^*, 2})$, respectively.
${\overline a}$ and ${\overline b}$ stand for those of
mirror chiral superfields
${\overline \Phi }({\bf 15^*, 1})$ and
${\overline \Phi }({\bf 6, 2})$, respectively.
Respective $Z_2$-charges ($R$-parity) of the superfields
are also listed.
\end{document}